\documentclass{desyproc}
\begin{document}
\def\greaterthansquiggle{\raise.3ex\hbox{$>$\kern-.75em\lower1ex\hbox{$\sim$}}}
\def\lessthansquiggle{\raise.3ex\hbox{$<$\kern-.75em\lower1ex\hbox{$\sim$}}}
\def\gl{\raise.3ex\hbox{$<$\kern-.68em\lower1ex\hbox{$>$}}}
\newcommand{\gts}{\greaterthansquiggle}
\newcommand{\lts}{\lessthansquiggle}
\newcommand{\cp}{\mbox{$\not \hspace{-0.15cm} C\!\!P \hspace{0.1cm}$}}
\newcommand{\comment}[1]{}

\title{Supersymmetry and New  Physics at $\gamma \gamma$ colliders\footnote{
Invited review  presented at the International Conference on the Structure
and the Interactions of the Photon including the 18th International Workshop
on Photon-Photon Collisions and the International Workshop on High Energy
Photon Linear Colliders, May 11-15, 2009.}
}

\author{{\slshape Rohini M. Godbole\\[1ex]
Centre for High Energy Physics, Indian Institute of Science, Bangalore, 
560012, India.\\ }}

\contribID{10}

\confID{1407}  
\desyproc{DESY-PROC-2009-03}
\acronym{PHOTON09} 
\doi  

\maketitle

\begin{abstract}
In this contribution we present a discussion of some aspects of the capabilities
of a photon collider to probe physics beyond the standard model. I will take
a few examples from Higgs physics, supersymmetry , extra dimensional theories
as well as unparticles, pointing out the special r\^ole that a photon collider
can play in each case.
\end{abstract}

\section{Introduction}
I have been asked to discuss new physics at the $\gamma \gamma$ 
collider~\cite{Badelek:2001xb}. 
In general, new physics can be discussed in two different ways:
\begin{itemize}
\item
[a)]In the framework of specific models proposed with a view to cure one 
or more of the ills of the SM, some well motivated and some speculative. 
Supersymmetry (SUSY), extra dimensional models (ED), little higgs models,  
noncommutative theories,unparticles etc. are some examples.
\item[b)]
The second is to look at the effect on different aspects
of phenomenology at a $\gamma \gamma$ collider
such as jet production, $t \bar t$ production or Higgs 
physics studies etc.,  in a  model independent manner.
\end{itemize}

In this contribution I  will pick some combination of the two above mentioned
strategies as well as that of the topics. I will  mainly concentrate on  
physics of the sparticles and (BSM) Higgs at the $\gamma \gamma$ colliders, 
trying to  identify  where the  $\gamma \gamma$ collider has a distinct 
advantage in terms of adding clarity to a  particular study, and/or 
increasing the coverage  in (SUSY) model parameter space as well as the reach 
in  masses, compared to the $e^+e^-$ option.
I will also include some discussion of new physics
such as extra dimensional theories or more speculative case of
unparticles in the context of $\gamma \gamma$ colliders.
\comment{Specific topics on which I have touched upon are:
a) 
Searching for  sparticles at $\gamma \gamma$ colliders {\it directly},
b) Study of the Heavy Higgs H/A sector and CP-mixing for 
CP-violating  (CPV) SUSY using the $WW/ZZ$ final states, $t/\tau$ polarisation
etc.,
c) SUSY parameter determination, and 
d) Probing new physics through its effects on the $\gamma \gamma$-Higgs
vertex or production of various final states such as dijets, pair of gauge 
bosons and/or fermion pairs.}

The two special features of the PLC  of great help in this are: 
the very accurate measurements ($\sim 2\%$) of the $\Gamma_{\gamma \gamma}$ 
decay width for the Higgs boson into two photons and good control on 
the polarisation of the
initial photon beams.

\section{SUSY: LHC Wedge, LEP hole and LHC/ILC}
It is necessary to summarise the LHC and LHC/ILC possibilities for SUSY 
studies and searches~\cite{Weiglein:2004hn}, before turning to a discussion
of the possibilities at the PLC. Recall that, the sparticle mass spectrum 
depends on the mechanism
responsible for SUSY breaking and can vary widely, {\it but} 
the sparticle  spins and couplings are predicted  unambiguously. To establish
SUSY with the help of the  two colliders  LHC and ILC, we need  to
find the sparticles,  measure their masses, spins and couplings.
Another thing is to note that for $\tilde\chi^\pm, \tilde\chi^0_l$ as well as
the supsesymmetric partners of the third generation of the quarks/leptons,
the masses as well as the couplings, can depend on the SUSY breaking
mechanism and parameters. The LHC will be able to 'see' the strongly interacting
sparticles if the SUSY scale is TeV. If the sparticle mass is within the 
kinematic reach of the ILC, we will be able to make accurate mass measurements
and also can make clean spin determination. In this situation the ILC can 
even help us determine the SUSY
model parameters and hence the SUSY breaking mechanism as has been summarised 
in the SPA program~\cite{AguilarSaavedra:2005pw}.  In spite of this very 
impressive and exhaustive coverage of SUSY by the LHC and the ILC in $e^+e^-$ 
mode,  there are a few 'holes' in the SUSY
parameter space. In the so called LHC wedge~\cite{AguilarSaavedra:2001rg}, 
$\tan \beta \simeq 4-10$, $M_A, M_H > 200$--$250$ GeV, only the light Higgs
$h$ of SUSY will be observable at the LHC and the $H/A$ will not be visible at
the first generation ILC. In case of CP violating MSSM also, there exists a
'hole' in the $\tan \beta$-$m_{H^\pm}$ plane, for low $\tan \beta~~\lts~ 3-5$.
This corresponds to three  neutral higgses $\phi_i, i=1-3$, which may not
be CP eigenstates, in the mass range 
$m_{\phi_1} < 50, 100 < m_{\phi_2} < 110$  and 
$130 < m_{\phi_3} < 180$ .  This region can not be ruled out by LEP 
searches and where the LHC also may not have reach~\cite{Accomando:2006ga}.
The $\gamma \gamma$ collider (PLC), can indeed offer unique possibilities in
this case. Further,  discovery of any charged scalar 
would uniquely signal physics beyond the SM (BSM).  In the 
following we will present examples of the special role that the PLC can play
in this context.

\section{Increased reach for new particle searches at the PLC} 
\begin{figure}[htb]
\centerline{
\includegraphics[width=0.40\textwidth]{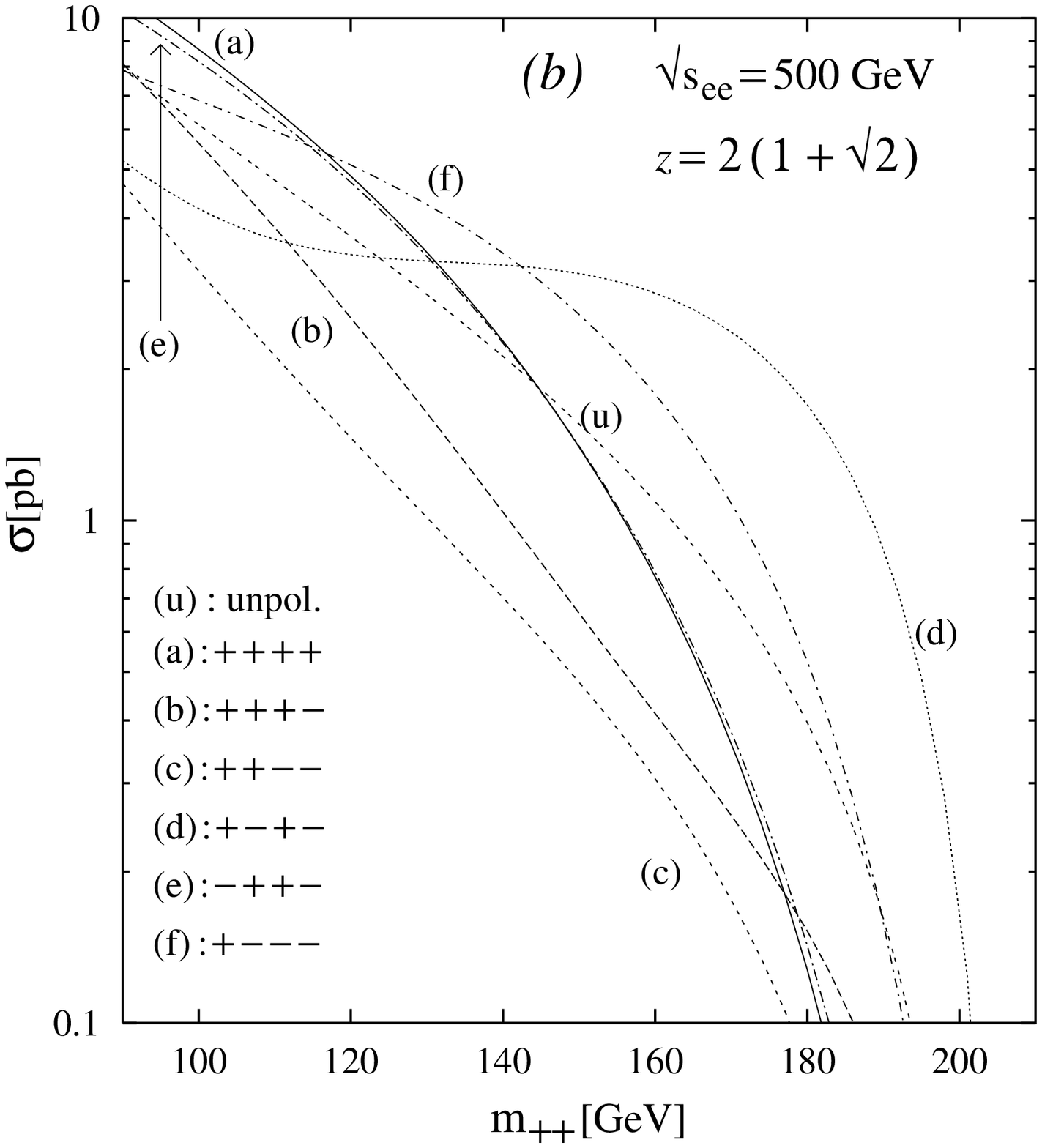}\hspace{0.5cm}
\includegraphics[width=0.55\textwidth]{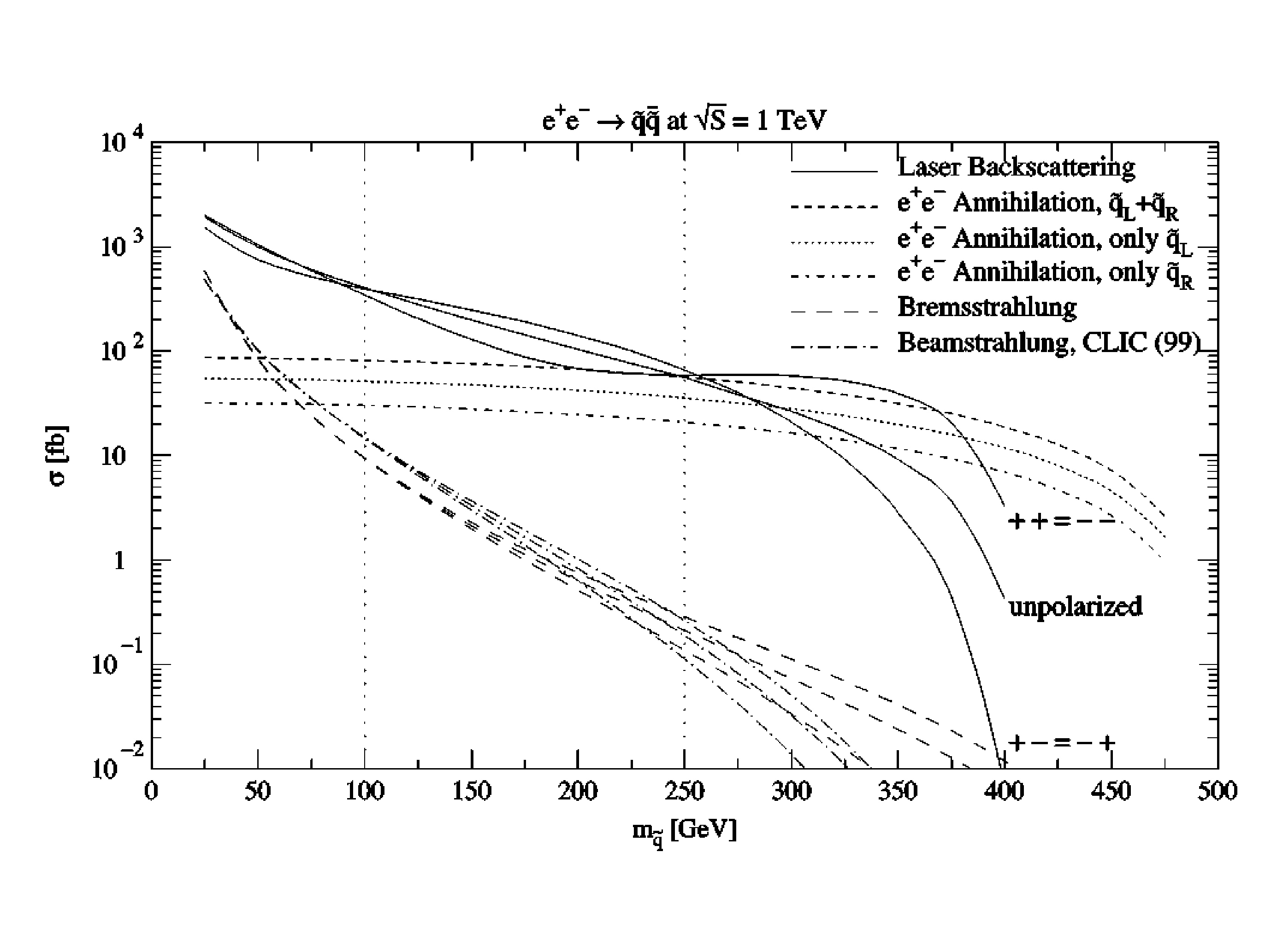}
}
\caption{Beam polarisation dependence of the production cross-section
charged scalars for doubly charged scalars(left 
panel)\protect~\cite{Chakrabarti:1998qy} and for squarks (right 
panel)\protect~\cite{Berge:2000cb} as a function of the  mass.}
\label{Fig1:godbole}
\end{figure}

As already mentioned charged Higgs, for that matter,
any charged scalar will be a signal of BSM physics beyond any doubt.
The production cross-sections of such scalars are enhanced in 
$\gamma \gamma$ collisions, compared to that in $e^+e^-$ collisions,
by a factor of $Q_S^2$, where $Q_S$ is the electromagnetic charge of the
scalar S. This is relevant, for example, in the little higgs models, which have
doubly charged scalars. Even for the singly charged scalars, the  dependence 
of the pair production cross-section on the original $e^\pm$ beam energy 
depends on the polarisation combination of the two beams and can be used to 
increase the mass reach. Right panel in Figure~\ref{Fig1:godbole}, taken from
Ref.~\cite{Chakrabarti:1998qy}, showing this polarisation dependence as 
function of $m_{H^{++}}$ illustrates this. In fact the right hand side panel of 
the same figure, taken from Ref.~\cite{Berge:2000cb},  
showing the cross-section for scalar quark production, both in
$e^+ e^-$ collisions and at the  $\gamma \gamma$ collider, as a function of
the scalar quark mass, directly illustrates how one can increase the reach 
in the charged scalar sector over the LC mode, at a $1$ TeV LC.

Above discussion clearly highlights the advantage offered by a PLC in 
case of charged scalars with a clear increase in the range of scalar masses
that can be probed with a given $e^+/e^-$ beam  energy.  An increase
in the range (by about a factor of $1.6$ ) in the reach in the mass of the 
heavy Higgs of SUSY (H/A),  due to the single Higgs production that is 
possible at the PLC, compared to that in pair production in $e^+e^-$ colliders, 
had also been noted in the context of MSSM higgs boson searches. This in 
fact can  fill  the LHC wedge region
of the MSSM parameter space~\cite{Muhlleitner:2001kw}.  In addition to this,
the PLC, in the $e$--$\gamma$ option can increase the range of 
$m_{\tilde e_R}$ mass if the mass
difference between the $\tilde e_R$ and $\tilde \chi_1^0$ is large. At  
an $e$-$\gamma$ collider,  the $e \gamma \rightarrow 
\tilde e_R \tilde \chi_1^\pm$ process has reach up to $m_{\tilde {e_R}} + m_{\tilde {\chi_1^0}} <  0.9 \sqrt{s_{ee}} $ , where as at an $e^+ e^-$ collider the
reach is $0.5 \sqrt{s_{ee}}$. This  has already been discussed in other talks
at this conference~\cite{Moenig:2009}.  Further, this is possible without the 
need of a polarised initial beam.  
\section{Better measurement of SUSY parameters}
Not just for the charged scalars but also for the new charged fermions, 
like $\tilde \chi_1^\pm$, pair production in $\gamma \gamma$ collisions, 
can  afford a good 
measurement of the B.R. $(\tilde \chi_1^\pm \rightarrow W \tilde \chi_1^0)$
and can  increase the accuracy of $\tan \beta $ 
determination in the SPA 
fit by over a factor $2$--$3$~\cite{Moenig:2009,Klamke:2005jj}, for a MSSM
point with parameter choice  very similar to that for  SPS 1a. 

In fact determination of $\tan \beta$ at an $e^+e^-$ collider is a particularly 
notorious for the lack of accuracy at large $\tan \beta$ in the process  
$\tilde\chi^+ \tilde \chi^-, \tilde\chi_j^0 \tilde \chi_i^0$ 
mainly due to the fact that  the observable involves 
$\cos 2 \beta$~\cite{Choi:2000ta}.
$\gamma \gamma \rightarrow \tau^+ \tau^- \phi \to \tau^+ \tau^- b \bar b$,
on the other hand  offers a  very good measurement of $\tan \beta$.  
Results of a phenomenological calculation~\cite{Choi:2004ne},
show that at $\tan \beta = 30$ it may be possible to have
$ \Delta \tan\beta = 0.9$--$1.3$, to be contrasted with an accuracy of about 
$10$--$12$~\cite{Choi:2000ta} at the $e^+e^-$ option. It is clear that this 
process has the potential to  help  enormously in SUSY parameter 
determination. However, this needs to be backed up by detailed simulations.
\section {Higgs physics and the PLC}
An accurate measurement of $\Gamma (\phi \rightarrow \gamma \gamma)$,
determination of the CP property of the Higgs, as well as measurement of CP
mixing in case of CP violation, are the three important ways in which a PLC
can make value addition compared to an $e^+ e^-$ collider. Many of 
these features, apart from the CP violation in the Higgs sector, both in 
the context of a particular model (SUSY) and in a model independent 
approach were already described in different 
talks~\cite{Moenig:2009,Spira:2009} at this conference. Almost any new 
physics, may it be SUSY, with and without CP 
violation~\cite{Djouadi:1998az,Belanger:2000tg,Moretti:2007th,Hesselbach:2007en}
or 2-higgs doublet model~\cite{Ginzburg:2001wj}  will in fact affect the
$\gamma \gamma$-Higgs couplings and hence the width. In the CPV MSSM or 
MSSM with non universal gaugino masses, these effects can be significant, 
yet being consistent with the current limits on all the sparticle masses.

A unique feature of a PLC is that the two photons can form a $J_z = 0$ state
with both even and odd CP. As a result, unlike the gauge boson fusion mode
which contributes mainly in the $e^+e^.-$ mode to the production,
the  PLC has a similar level of
sensitivity for both the CP-odd and CP-even components of a CP-mixed state:
\begin{equation}
  {\rm CP\!-\!even:}
  \epsilon_1\cdot \epsilon_2 = -(1+\lambda_1\lambda_2)/2 , \quad
  {\rm CP\!-\!odd:}
  [\epsilon_1 \times \epsilon_2] \cdot k_{\gamma}
  =\omega_{\gamma} i \lambda_1(1+\lambda_1\lambda_2)/2,
\end{equation}
$\omega_i$ and $\lambda_i$ denoting the energies and  helicities of the
two photons respectively; the helicity of the system is equal to
$\lambda_1-\lambda_2$.
This contrasts the $e^+e^-$ case, where it is possible  to discriminate between
CP-even and CP-odd particles but may be difficult to detect small CP-violation
effects for a dominantly CP-even Higgs 
boson~\cite{Accomando:2006ga,Godbole:2004xe}.

In this talk I should like to concentrate on the CP violation and anomalous 
$hVV$ couplings in the Higgs sector pointing out the role that the PLC can 
play in their study, after briefly mentioning the prominent issues in the 
first two topics.
\subsection{LHC-wedge}
At large $\tan \beta$ region, the dominant decay mode of the $H/A$ is into the
$b \bar b$ channel, where the $b \bar b$ background can be controlled by a
judicious choice of photon polarisation. $H/A$ separation can be achieved 
by choosing  polarisation vectors of the two photons to be perpendicular 
and parallel;  but this has implications for the QED $q \bar q$ background as 
well. Results of 
a detailed simulation~\cite{Spira:2006aa}, which were already discussed at 
this conference, show that  for a light higgs  it would be possible to measure
the $\gamma \gamma$ rates accurate to $\simeq 2 \%$ whereas for  $H/A$ 
measurement precision would be somewhat worse : $\sim 11\%$--$21\%$ .
In fact in these region the Supersymmetric decay of the $H/A$ into 
$\tilde \chi^\pm ,\tilde \chi^0_l$ pairs can also  be 
used~\cite{Muhlleitner:2001kw}.
\subsection{CP properties and CP violation in the Higgs sector}
In the MSSM the properties of the  Higgs sector, at the tree level,
are determined in terms of two parameters $\tan \beta$ and $\mu$.  If some of
the SUSY parameters have nonzero phases, then the Higgs sector can have loop
generated CP violation, even with a CP conserving tree level scalar 
potential~\cite{Pilaftsis:1999qt}. Recall the existence of the 'LEP-hole'
mentioned earlier. Effect  of this CP violation on the masses 
and the coupling of the Higgses in this parameter range, can also affect the 
LHC reach and part of the 'hole' remains~\cite{Accomando:2006ga}, even after 
the recovery of some part of the parameter space through the decay chain 
$t \rightarrow b H^+ \rightarrow b W h \rightarrow b W b \bar b$~\cite{Ghosh:2004cc}. 

A PLC will be  able to produce such a neutral Higgs in all cases; 
independent of whether it is a state with  even/odd or indeterminate CP parity.
For the PLC, one can form three polarization asymmetries in terms of helicity
amplitudes which give a clear measure of CP mixing~\cite{Grzadkowski:1992sa}.
Note however that these require linearly polarised photons in addition to the 
circular polarisation.  With circular beam polarization almost mass 
degenerate (CP-odd) $A$ and (CP-even) $H$ of the MSSM may be separated
\cite{Muhlleitner:2001kw,Spira:2006aa}.
In addition, one can use information on the 
decay products of $WW$, $ZZ$~\cite{Niezurawski:2004ga}.
Further, Higgs contribution to $\gamma \gamma \rightarrow f \bar f$
can give nontrivial information on the CP 
mixing~\cite{Asakawa:1999gz,Godbole:2002qu,Asakawa:2003dh,Ellis:2004hw,Choi:2004kq,Godbole:2006eb,Accomando:2006ga}.

\newpage

\noindent$\bullet$ \underline{\bf $f \bar f$ final state.}

\vspace{0.1cm}

The process 
receives contributions from the $s$--channel Higgs exchange and the 
$t$-channel QED diagram.
It is possible to determine the CP mixing, if present, by using the 
polarisation of the initial state $\gamma$ or that of the fermions into 
which the  $\phi_i$  decays.
In MSSM the CP-even $H$ and the CP-odd $A$ are degenerate.
In  the situation that the mass
difference between the two is less than the sum of their widths, a coupled 
channel analysis technique~\cite{Pilaftsis:1997dr} has to be used. The authors 
of Refs.~\cite{Ellis:2004hw} and \cite{Choi:2004kq} explore the use of
beam polarisation and final state fermion polarisation to analyse
this situation whereas the use of decay fermion polarisation for 
determination of the Higgs CP property for a generic choice of the MSSM 
parameters is explored in Ref.~\cite{Godbole:2006eb}.

The most general couplings of a Higgs to $f \bar f$ and $\gamma \gamma$ can
be written in a model independent way, accounting for possible CP violation,
as~\cite{Godbole:2002qu,Asakawa:2003dh}:
\begin{eqnarray*}
{\cal V}_{f \bar f\phi}&=&-ie\frac{m_f}{M_W} \left(S_f+ i\gamma_5P_f\right),\\
{\cal V}_{\gamma\gamma\phi}&=&\frac{-i\sqrt{s}\alpha}{4\pi}\left[S_{\gamma}(s)
\left(\epsilon_1.\epsilon_2-\frac{2}{s}(\epsilon_1.k_2)(\epsilon_2.k_1) \right)
\right.
-\left. P_{\gamma}(s)\frac{2}{s}\epsilon_{\mu\nu\alpha\beta}\epsilon_1^{\mu}
\epsilon_2^{\nu} k_1^{\alpha} k_2^{\beta} \right].
\end{eqnarray*}
When we consider this in the context of a particular model then the 
form-factors,
$\{S_f,P_f,S_{\gamma},P_{\gamma}\}$ depend upon model parameters. For example,
for the CP violating MSSM these depend on $m_{H^+}, \ \tan\beta,
 \ \mu, \ A_{t,b,\tau}$, $\Phi_{t,b,\tau}, \ M_{\tilde q}, \ M_{\tilde l}$
etc. 
\begin{figure}[htb]
\centerline{
\includegraphics[width=6cm,height=5.0cm]{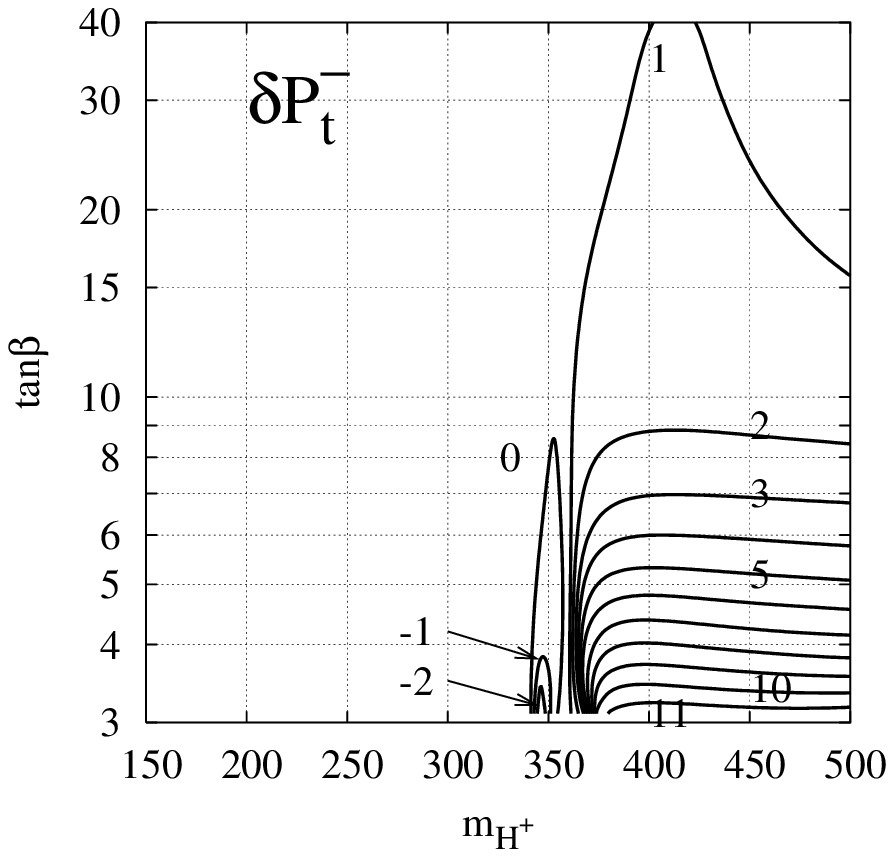}\hspace{0.5cm}
\includegraphics[width=6cm,height=5.0cm]{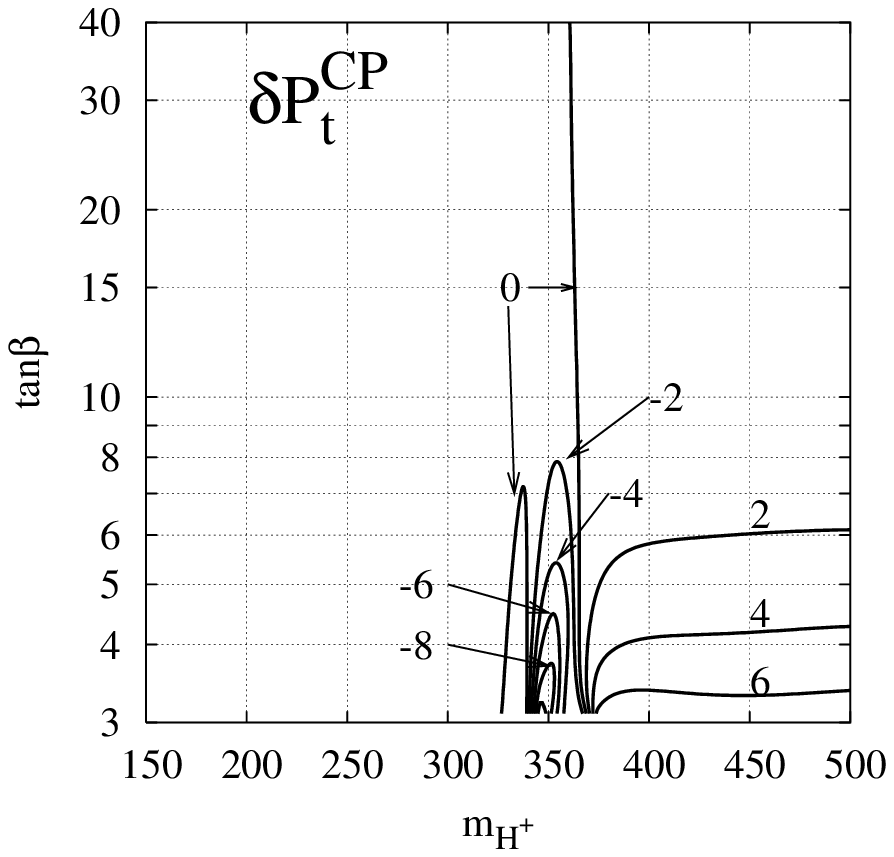}
}
\caption{Expected values (in $\%$) of 
$\delta P_t^-$ and $\delta P_t^{CP}$ for the
top quarks produced in the process, $\gamma \gamma \rightarrow t \bar t$,
including the $s$-channel Higgs exchange contribution, in the CPV MSSM,
in the CPX scenario, in the $m_H^+$--$\tan \beta$
plane\protect~\cite{Godbole:2006eb}}.\label{Fig3:godbole}
\end{figure}
The model independent case and the specific case of CP violating SUSY
are analysed in  Refs.~\cite{Godbole:2002qu,Asakawa:2003dh} 
and ~\cite{Godbole:2006eb} respectively.
The helicity amplitude for the production will in general involve CP 
even combinations such as $S_f \Re(S_\gamma) ({\rm viz.} x_i)$   as well 
as CP odd-combinations such as $S_f \Im(P_\gamma) ({\rm viz.} y_i)$.
Note that the QED background is  $P$, $CP$ and chirality conserving.
Higgs exchange diagram violates these symmetries.  This means that
in the presence of the Higgs, existence of chirality flipping interaction
imply nonzero values of  the various $\{x_i,y_j\}$ which in turn means that the 
fermion-polarisation carries a footprint of the Higgs contribution as well as
any CP violation in the $\phi \gamma \gamma$ and $\phi f \bar f$ couplings.
It is also very gratifying that the heavier fermions $t,\tau$ 
which have the largest $\phi f \bar f$ coupling are also the fermions
whose polarisation is amenable to experimental measurements.
The polarisation of the initial state $\gamma$
can be controlled by adjusting the initial laser and the $e$ polarisation. The
$\phi$ contribution is enhanced using the combination $\lambda_e \times 
\lambda_l = -1$. One can construct observables, with unpolarised and polarised
laser and $e$ beams in terms of expected fermion polarisation: $P_f^{U}$ being 
the expected one for unpolarised initial states  and $P_f^{++}, P_f^{--}$  being
the observables with polarised beams. Here $+/-$ in the (double) superscripts 
refer to the polarisation of the $e, \lambda_e$. $P_f^{++}, P_f^{--}$ are 
nonzero even for the QED diagrams alone, but the $P$ invariance of QED implies
$ P_f^{++} = -P_f^{--}$ Hence a nonzero value for $ P_f^{++} +P_f^{--}$
will clearly indicate parity violation.  In case of $C$ invariance this then 
is also CP violating. Thus $P_f^{U}$ and 
$\delta P_f^{CP} = P^{++}_f + P^{--}_f$ are both probes  of CP violating 
contribution. Further, 
$P_f^{++}$ is  modified by the Higgs contribution such that 
$P_f^{++}-(P_f^{++})^{QED}\neq 0$ even if $\phi$ is $CP$ eigenstate. Hence,  
$\delta P_f^+ = P^{++}_f - (P^{++}_f)^{QED}$ and $\delta P_f^- = 
P^{--}_f - (P^{--}_f)^{QED}$  are both probes of chirality flipping
interactions.  Figure ~\ref{Fig3:godbole}, taken from 
Ref.~\cite{Godbole:2006eb}, shows the values for $\delta P_t^-$ 
and $\delta P_t^{CP}$, expected in the CP-violating MSSM, 
in the  $m_H^+$--$\tan \beta$ plane, for the CPX 
scenario~\cite{Pilaftsis:1999qt}, in the 
left and the right panel respectively. 
\begin{figure}[ht]
\centerline{
\includegraphics[scale=0.48]{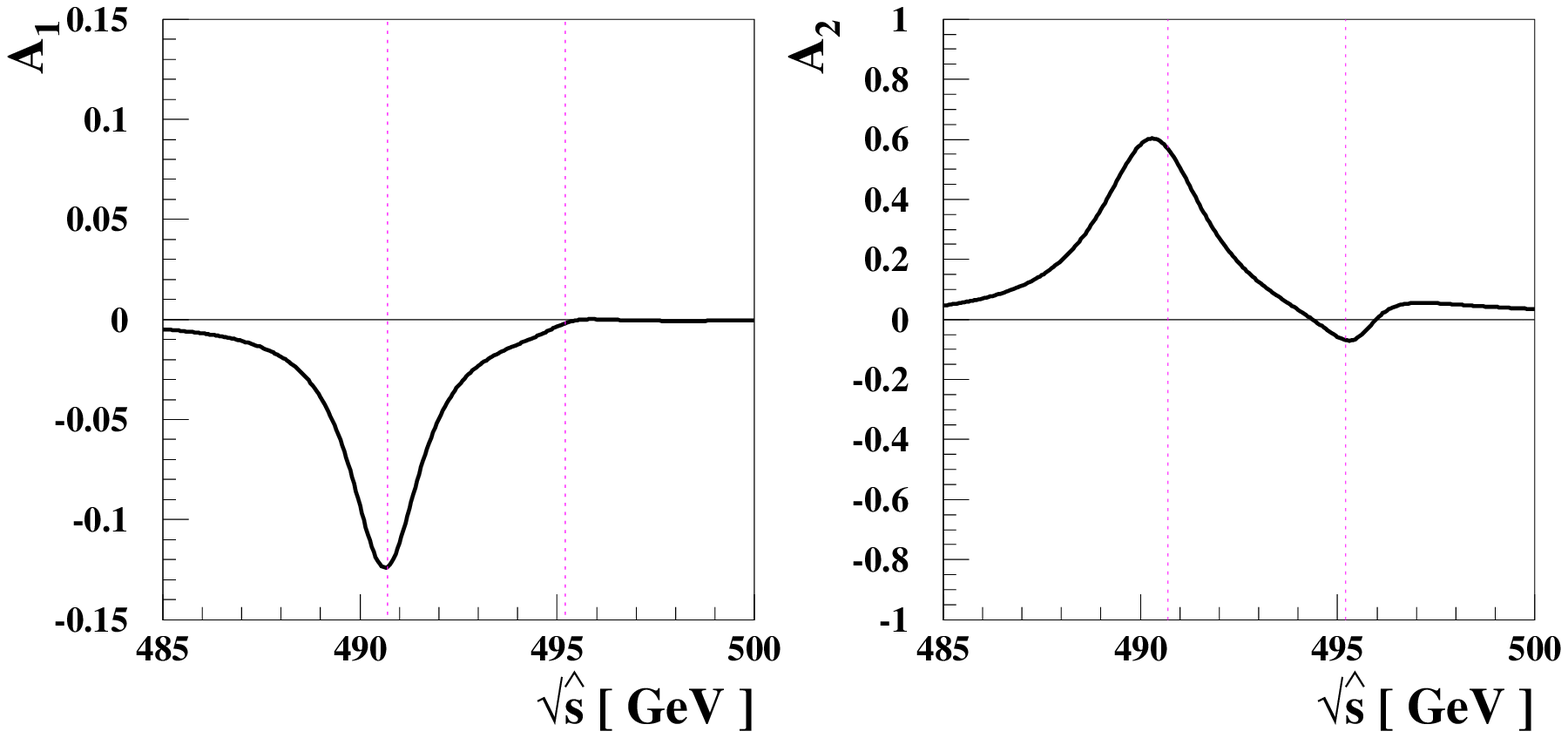}\hspace{0.1cm}
\includegraphics[scale=0.52]{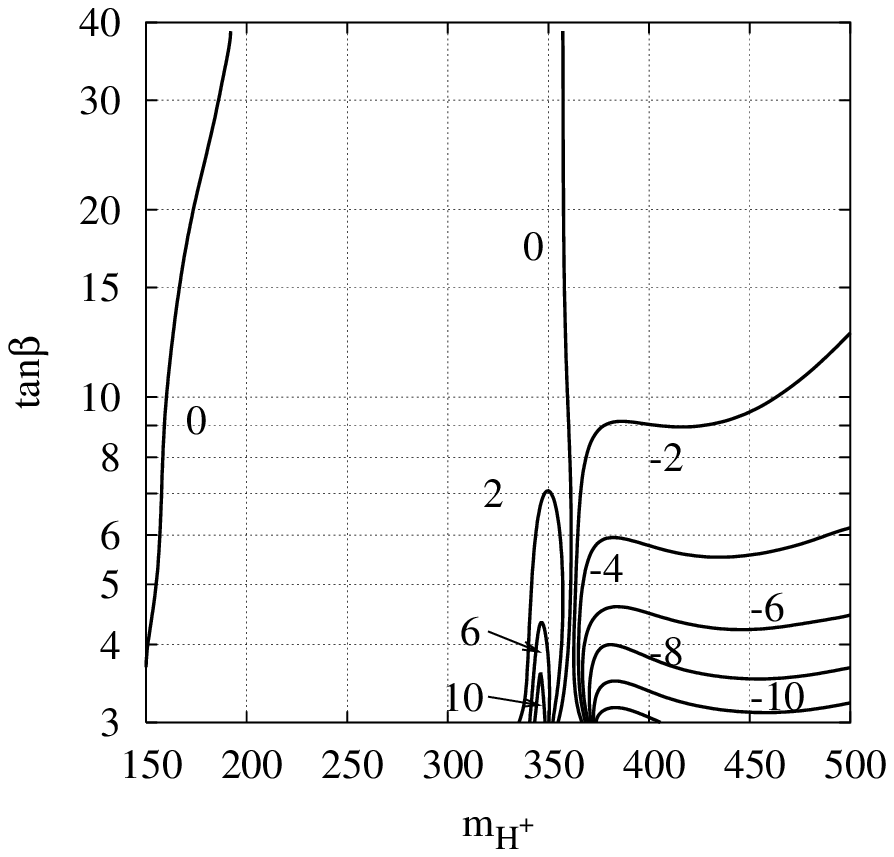}
}
\caption{CP violating asymmetries constructed out of different
combination of cross-sections with final state $t$ helicity and initial
state photon helicity for top quarks produced in the process, 
$\gamma \gamma \rightarrow t \bar t$,
including the $s$-channel Higgs exchange contribution, in the 
degenerate $\phi_2/\phi_3$ case in CPV MSSM as a function
of the $\gamma \gamma$ centre of mass energy (two left 
panels)~\cite{Ellis:2004hw} and the mixed lepton charge-photon 
helicity asymmetry (in$\%$) for the generic case in the CPX 
scenario, in the $m_H^+$--$\tan \beta$ plane with a choice of $e^+e^-$ beam 
energy to maximise the asymmetries (right panel)\protect~\cite{Godbole:2006eb}.}\label{Fig6:godbole}
\end{figure}   
A similar calculation of the expected $\tau$ polarisation indicates that the two
fermion polarisation offer coverage in the complementary regions of this
parameter space and part of the 'LEP'-hole, which can not be covered at the LHC,
can in fact be covered by these measurements. Note that the ILC can provide 
part coverage of the region through production of other Higgs bosons, but 
$\gamma \gamma$ collisions and/or production in the decay of the charged 
higgs~\cite{Ghosh:2004cc}, remain the only two modes for the light neutral 
higgs, in this left over region of the 'LEP-hole'.

The expected values of the various observables presented in the 
Figure~\ref{Fig3:godbole} and the rightmost panel of~\ref{Fig6:godbole}, 
are for a value of the common CP violating phase
$\Phi = 90^\circ$. and the beam energy is adjusted for
each point in the  scan such that the peak of the photon spectrum matches with
scaled mass $ m_{\phi}/\sqrt{s_{\gamma \gamma}} $, $m_\phi$ being the average
mass of the two states which may be close in mass.
Nowhere in this range of the parameters are such that  the
two states are extremely degenerate, and hence a coupled channel analysis is
not required.  We see that even in this case, the size of the expected
asymmetries is not too small. Thus in a generic case of CPV MSSM, the PLC
can probe this CP-mixing in the Higgs sector.
In case of extreme degeneracy of the two states, the expected polarisation
asymmetries for both the fermion final states containing  $\tau$ and $t$, 
are enhanced resonantly and the observability is increased even more.
The left two  panels of Figure~\ref{Fig6:godbole} taken from 
Ref.~\cite{Ellis:2004hw} show two such 
CP violating asymmetries constructed out of combination of cross-sections
with different helicities for the initial state photons  and final state 
quarks, as a function of $\sqrt{s_{\gamma \gamma}}$, with model parameter
values chosen to maximise the effect. 

Since the top quark decays before it hadrnoises, the  decay products retain
the  top quark spin information. In fact, the decay lepton
angular distribution is a particularly good probe of this polarisation
due to the independence of the correlation between the polarisation and the 
angular distribution and  any anomalous $tbW$ vertex~\cite{Godbole:2006tq}.
The above mentioned polarisation asymmetries translate into 'mixed 
beam polarisation-lepton charge asymmetry' constructed out of cross-sections
with different photon helicity combinations as in 
Ref.~\cite{Godbole:2002qu}, but for values of the form factors 
$S_f,P_f,S_{\gamma},P_{\gamma}$ , calculated in the CPX scenario
as a function of the MSSM parameters $\tan\beta$--$m_{H^+}$ plane.
These, taken from Ref.~\cite{Godbole:2006eb} are shown  in the rightmost
panel of the Figure~\ref{Fig6:godbole}.

\noindent$\bullet$ \underline{\bf 2HDM and $WW/ZZ$ final states at a PLC}

\vspace{0.1cm}

As mentioned earlier, CP violation in the Higgs sector has also been studied
in the context of a two Higgs doublet model and the possibility of determining
the CP violating phase through the kinematic distribution of the decay products
of the $W$ and the $Z$~\cite{Niezurawski:2004ga}, with a realistic photon 
spectra has been investigated. The phase $\Phi_{CP}$ 
and the relative strength of the $\phi VV$ coupling relative to that
in the SM can be measured to about $\lts 0.02$--$0.05$ depending on the
mass of the Higgs.. The errors are 
computed, assuming the SM value of  $0$ and $1$ respectively for the two.  
The interesting part of this study is the fact that the two photon width,
its phase and the relative normalisation of both samples, are all allowed 
to vary in the fit. The former, which is available only at a PLC,
is seen to impact the results significantly.

In fact in the $e \gamma$ option  the photon collider offers also a unique 
possibility of determining accurately the $hWW$ anomalous coupling. The
accuracy of determination of this coupling in the $e^+e^-$  option is
limited due to the big background from the $ZZh$ contribution to the
same final state. This does not require polarised beams 
either~\cite{Choudhury:2007zz}.
\subsection{Higgs self coupling and the $\gamma \gamma$ collider}
The ILC in the $e^+e^-$ mode offers only a very limited information on
the trilinear $hhh$  coupling~\cite{Weiglein:2004hn,Djouadi:2007ik}. This 
information can be obtained through a study of Higgs pair production at
a $\gamma \gamma$ collider and is shown to  be a good probe of $hhh$ couplings 
and comparable perhaps to other probes at the LHC and
the ILC. Recently, the modification of the $hhh$ coupling in the framework
of a general two Higgs doublet models was addressed in a couple of 
analyses~\cite{Asakawa:2008se,Arhrib:2009gg,Cornet:2008nq}.
If the modification of the  $hhh$ coupling is due to new particles
in the spectrum, then it will also modify  $h \gamma \gamma $
as well the $\gamma \gamma h  h$ coupling. So in the framework of a
two higgs doublet model they calculate the net change in the cross-section
$\gamma \gamma \rightarrow hh$ and show that the sensitivities possible 
at the PLC can indeed test these models.
\section{Extra Dimensional models and the PLC}
In the context of models with TeV scale gravity; ie. models with
extra dimensions, the $\gamma \gamma$ production of  all the matter and gauge
boson fields is  altered substantially. The extra dimensions can be
probed in the dijet final state~\cite{Ghosh:1999ex}, through the gauge boson
couplings to a pair of photons~\cite{Rizzo:1999sy}, in the production of a
$t \bar t$ pair~\cite{Mathews:1999ik} as well as in the
$e \gamma$ mode~\cite{Ghosh:1999dg}, up to a scale comparable and/or somewhat
higher level, compared to the LC option. However, all these calculations 
have been done at the theorists level and an evaluation of the net gain due to 
PLC, when a realistic photon spectrum is used, is not available.
\section{Unparticles and the PLC}
\begin{figure}[ht]
\centerline{
\includegraphics[scale=0.25,angle=-90]{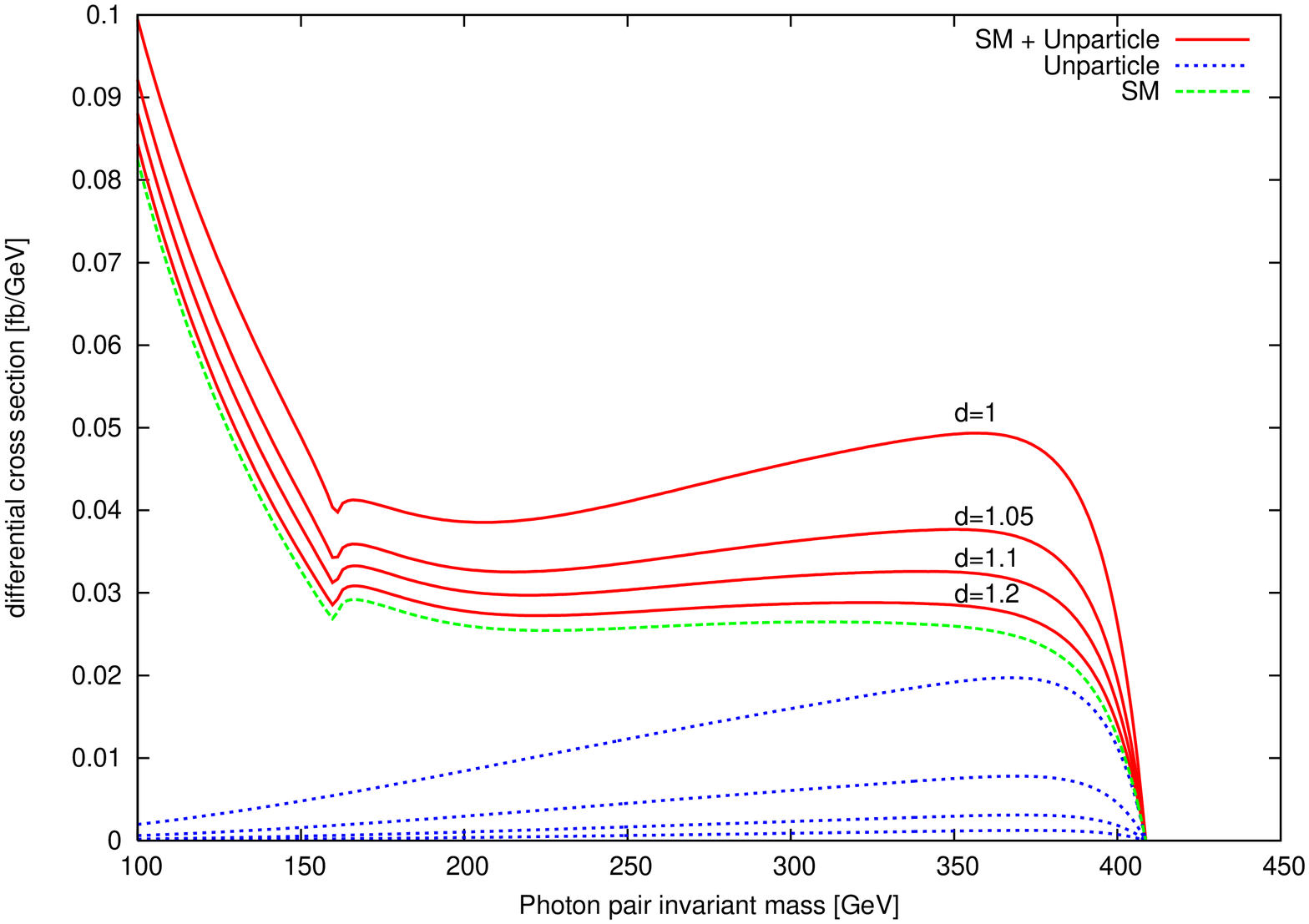}\hspace{0.5cm}
\includegraphics[scale=0.25,angle=-90]{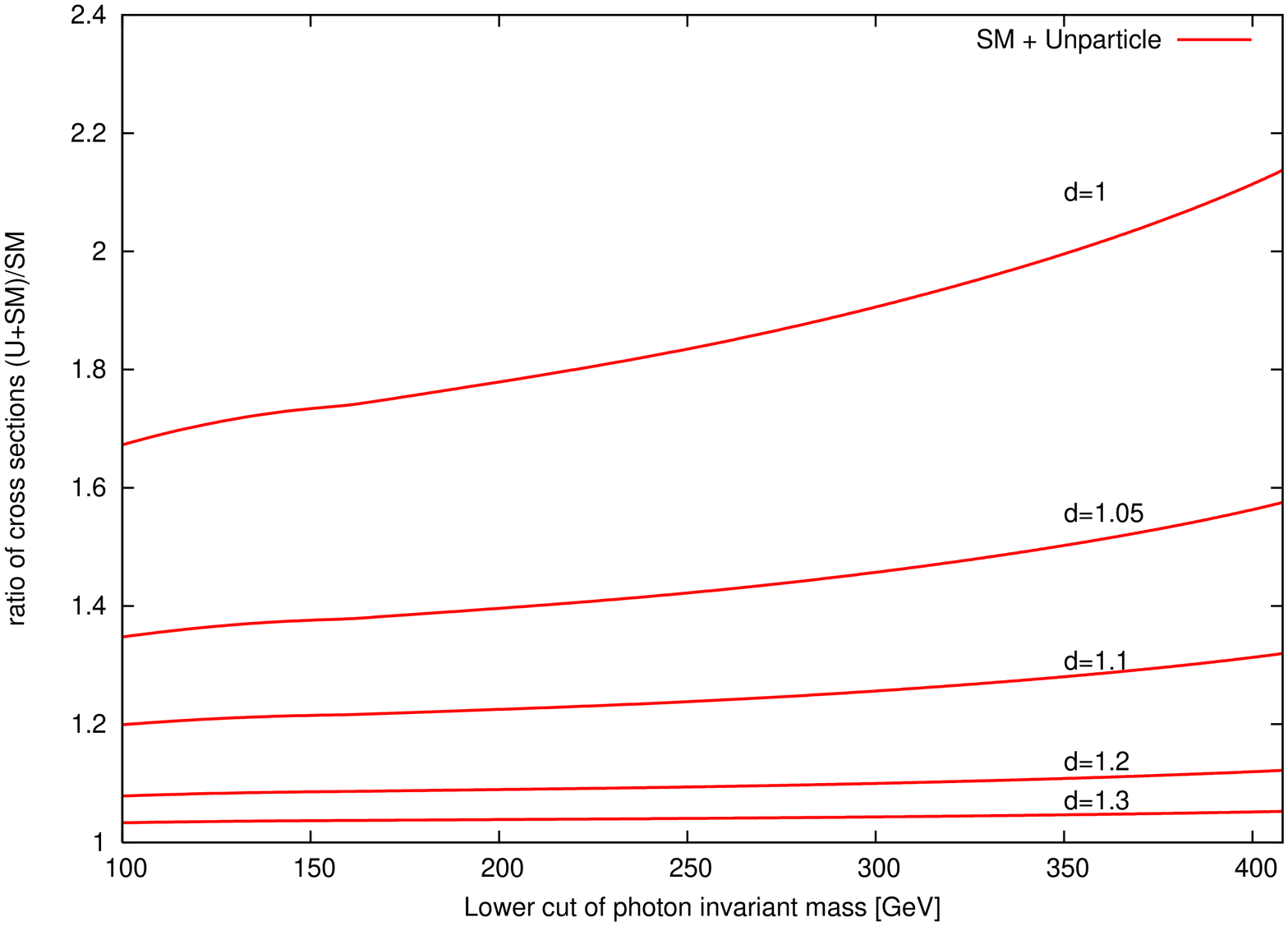}
}
\caption{The $\gamma \gamma$ invariant mass distribution for a 500 GeV
machine showing the effect of scaling dimension of the unparticles(left panel)
and the  values of $\sigma_{\cal U} + \sigma_{SM} / \sigma_{SM}$ as
a function of $\gamma \gamma$ centre of mass energy(right 
panel)~\protect\cite{Kikuchi:2008pr}.}
\label{Fig5:godbole}
\end{figure}
Along with the very well motivated physics beyond the SM like Supersymmetry 
and extra dimensions,
the PLC can also probe  speculative physics like unparticles. Among the 
different discussions that exist, I am going to give only one example
taken from~\cite{Kikuchi:2008pr}, where they consider the effect of 
unparticles on the process $\gamma \gamma \rightarrow \gamma \gamma$,
which can be studied at a photon collider. 
In these theories a  hidden conformal sector 
provides ``unparticle'' which couples to the Standard Model sector through 
higher dimensional operators in low energy effective theory. If one focuses on
operators which involve  unparticle, the Higgs doublet and the gauge bosons,
after the Electroweak Symmetry breaking, a mixing between unparticle and 
Higgs boson ensues.  In turn this can cause sizable shifts for the couplings 
between Higgs boson and a pair of photons~\cite{Kikuchi:2007qd}. 
Since the process proceeds in the 
SM only at loop level, it has a great potential to probe new physics.
The authors of Ref.~\cite{Kikuchi:2008pr}  show that $\gamma \gamma$ 
collider in this case can be sensitive to a scale of 5 TeV for $\sqrt{s} = 
500$ GeV. The plot in the left panel of Figure~\ref{Fig5:godbole} 
 taken from this reference, shows the  cross-section as a function of 
$\gamma \gamma$ invariant mass. The structure in this
distribution  reflects the scaling dimension of unparticle.
The second plot (right panel) in Figure~\ref{Fig5:godbole}  
shows ratio of $\sigma_{\cal U} + \sigma_{SM} / \sigma_{SM}$.
\section{Conclusions}
Thus  a PLC can play an important and unique role 
in many ways in probing BSM physics.   
Loop effects on $\gamma \gamma$ processes and couplings can probe
it indirectly. 
Further, it can affect search
prospects  of new charged scalars, a sure harbinger of New Physics,
by providing comparable  reach, if not more, as the  
$e^+e^-$ option for a TeV energy LC. Polarisation dependence of
the photon spectrum and cross-section can play an important role.
 $\Delta \beta \simeq 1$ at large $\tan \beta$  can be achieved using
$\tau \tau$ fusion. There are major gains for the SUSY Higgs sector as it 
provides reach for $H/A$ in regions where LHC does not have any. The $s$ channel
production increases  reach in the mass of  neutral Higgses by a factor 
$\sim 1.6$ due to single production that is possible.
 Advantages of a $\gamma \gamma$ collider are even more if
CP violation is present in the Higgs sector. The polarisation asymmetries 
constructed using initial state photon polarisation and final state fermion 
polarisations, can be a very good probe of the CP violation in the Higgs 
sector.
The $H/A$ contribution can be probed therefore through mixed
polarisation-charge asymmetries, i.e asymmetries in initial state polarisation
and final state lepton charge. If CP violation makes the lightest higgs
dominantly pseduoscalar and hence 'invisible' at LEP/ILC/LHC, then
$\gamma \gamma$ collider is the only place it can be produced directly.
The PLC is capable of probing new physics such as extra dimension through
production of dijets, top pairs, gauge bosons etc. in $\gamma \gamma$ 
collisions. 
\section{Acknowledgments}
I wish to complement the organisers for  the excellent organisation of 
this meeting.
This work was partially supported by the Department of Science and Technology,
India under the grant of the J.C. Bose Fellowship, under project number
Grant No. SR/S2/JCB-64/2007.
 

\begin{footnotesize}
\bibliographystyle{unsrt}
\bibliography{gr_submit}
%
\end{footnotesize}


\end{document}